\documentclass[twocolumn]{aastex62}

\usepackage{graphicx}
\newcommand\teff{$T_{\mathrm{eff}}$}
\newcommand\teffq{$T_{\mathrm{eff}}^4$}
\newcommand{\gf}{$g_{F}$} 
\newcommand\logg{$\log\,g$}
\newcommand\loggf{$\log g_{F}$}

\newcommand\teffeq{T_{\mathrm{eff}}}

\newcommand\mbol{$m_{\mathrm{bol}}$}
\newcommand\Mbol{$M_{\mathrm{bol}}$}
\newcommand\Msun{$M_{\odot}$}
\newcommand{\lgf}{$\log\,g_{\!\mbox{\scriptsize \,\sc f}}$} 
\submitjournal{ApJ}

\shorttitle{The extended FGLR}
\shortauthors{Kudritzki et al.}

\begin{document}

\title{A Simple Unified  Spectroscopic Indicator of Stellar Luminosity: the Extended Flux Weighted Gravity - Luminosity Relationship}

\correspondingauthor{Rolf-Peter Kudritzki}
\email{kud@ifa.hawaii.edu}

\author{Rolf-Peter Kudritzki}
\affiliation{LMU M\"unchen, Universit\"atssternwarte, Scheinerstr. 1, 81679 M\"unchen, Germany}
\affiliation{Institute for Astronomy, University of Hawaii at Manoa, 2680 Woodlawn Drive, Honolulu, HI 96822, USA}
\author{Miguel A.\ Urbaneja}
\affiliation{Institut f\"ur Astro- und Teilchenphysik, Universit\"at Innsbruck, Technikerstr. 25/8, 6020 Innsbruck, Austria}
\author{Hans-Walter Rix}
\affiliation{Max Planck Institute for Astronomy, K\"onigstuhl 17, 69117 Heidelberg, Germany}
%\author{Eva Sextl}
%\affiliation{LMU M\"unchen, Universit\"atssternwarte, Scheinerstr. 1, 81679 M\"unchen, Germany}
\

\begin{abstract}
We show that for a wide range of stellar masses, from 0.3 to 20 \Msun, and for evolutionary phases from the main sequence to the beginning of the red giant stage, the stellar flux weighted gravity, \gf $\equiv$ g/\teffq, is tightly correlated with absolute bolometric magnitude \Mbol. Such a correlation is predicted by stellar evolution theory. We confirm this relation observationally, using a sample of 445 stars with precise stellar parameters. It holds over 17 stellar magnitudes from \Mbol~= 9.0~mag to -8.0~mag with a scatter of 0.17 mag above \Mbol~= -3.0 and 0.29 mag below this value. We then test the relation with 2.2 million stars with 6.5 mag $\ge$ \Mbol~ $\ge$ 0.5 mag, where 'mass-produced' but robust \logg, \teff~and \Mbol~from LAMOST DR5 and Gaia DR2 are available. We find that the same relation holds with a scatter of $\sim$0.2~mag for single stars offering a simple spectroscopic distance estimate good to $\sim$ 10\%.
\end{abstract}


\section{Introduction}

With the seminal work by Annie Jump Cannon and her collaborators at Harvard Observatory \citep{Cannon1918a, Cannon1918b} it has become clear that the spectra of stars contain information about their intrinsic brightness. Since then it is a dream of stellar astronomers to develop a simple way to estimate absolute stellar magnitudes from stellar spectra and to use them for distance determinations. This has led to the methodology of spectroscopic parallaxes where classification of spectral types and luminosity classes in conjunction with calibrated values of absolute magnitudes is used for distance determinations. Over the years and with large amounts of data available for calibration this method has become more and more sophisticated and precise, in particular, when restricted to stellar subtypes (see, for instance, \citealt{Coronado2018, Hogg2019}).

In this paper we pursue a somewhat different direction. We introduce a very simple unified spectroscopic method to estimate absolute stellar magnitudes. It is directly motivated by stellar physics and applies to stars in a wide range of stellar masses, between 0.3 to 20 \Msun, in all evolutionary phases from the ZAMS to the beginning of the red giant phase and is surprisingly precise. The method makes use of the flux weighted gravity, \gf $\equiv$ g/\teffq, which can be straightforwardly derived from stellar spectra by using model atmosphere methods, which are now standard and widely distributed. The fact that \gf~ is tighly correlated with absolute stellar magnitude \Mbol~in the case of massive blue supergiant stars was first realized by \cite{Kudritzki2003, Kudritzki2008}, who discovered the flux weighted gravity - luminosity relationship (``FGLR''), which was then subsequently used for extragalactic distance determinations (see \citealt{Kudritzki2016} and references therein). \cite{Langer2014} noted the great potential of the ``Spectroscopic Hertzsprung-Russel Diagram (sHRD)'', where \gf~ is used instead of luminosity to discuss stellar evolution, and \cite{Anderson2016} discovered that Cepheid stars also follow a FGLR. As will be shown in the following sections, the use of flux weighted gravity as an indicator is not restricted to the most massive stars, but can also be extended to stars of much lower mass.

%\begin{figure*}[htp!]
 \begin{figure*}[ht!]
  \begin{center}
    \includegraphics[scale=0.40]{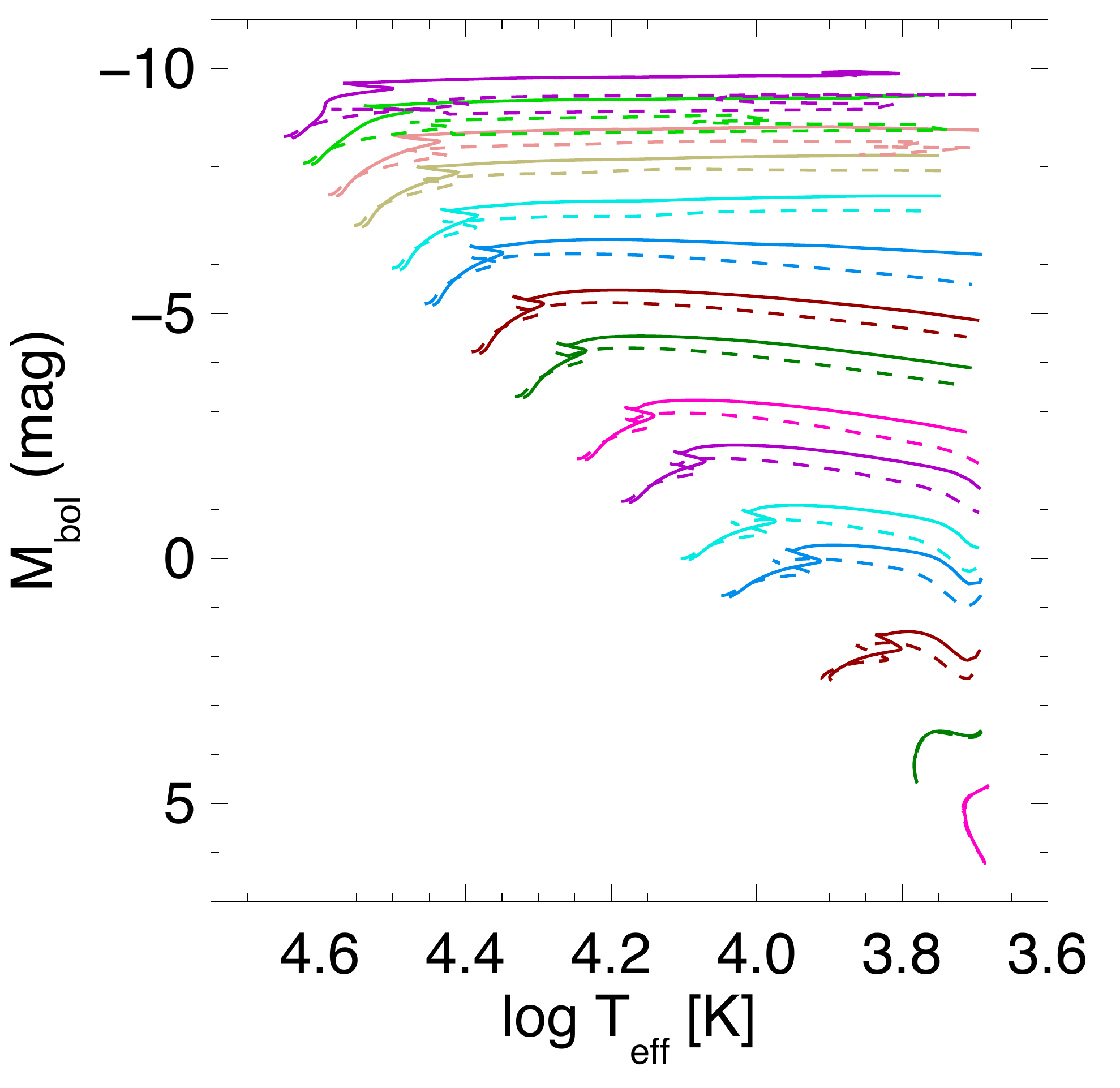}
    \includegraphics[scale=0.40]{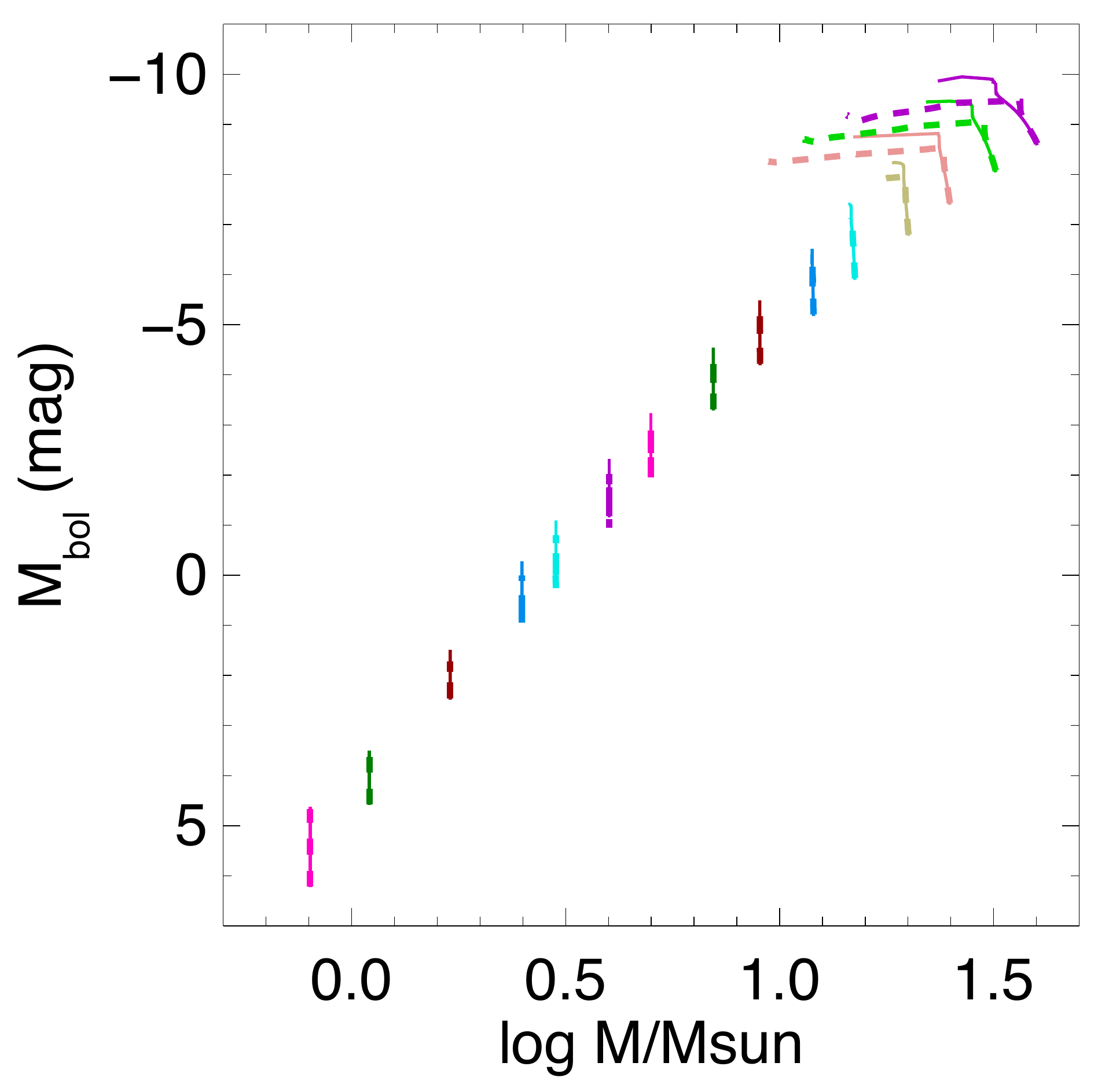}
    \includegraphics[scale=0.40]{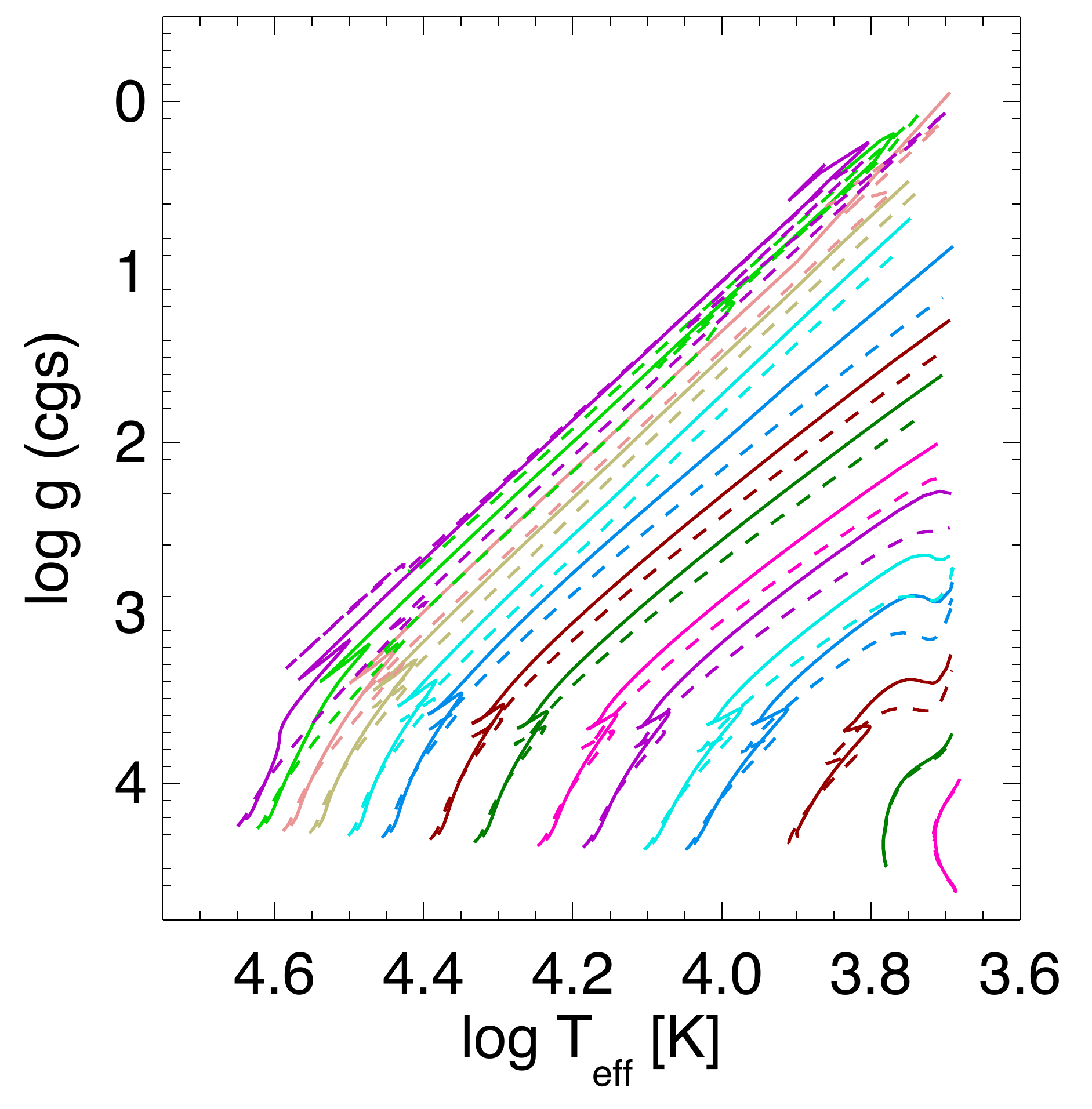}  
    \includegraphics[scale=0.40]{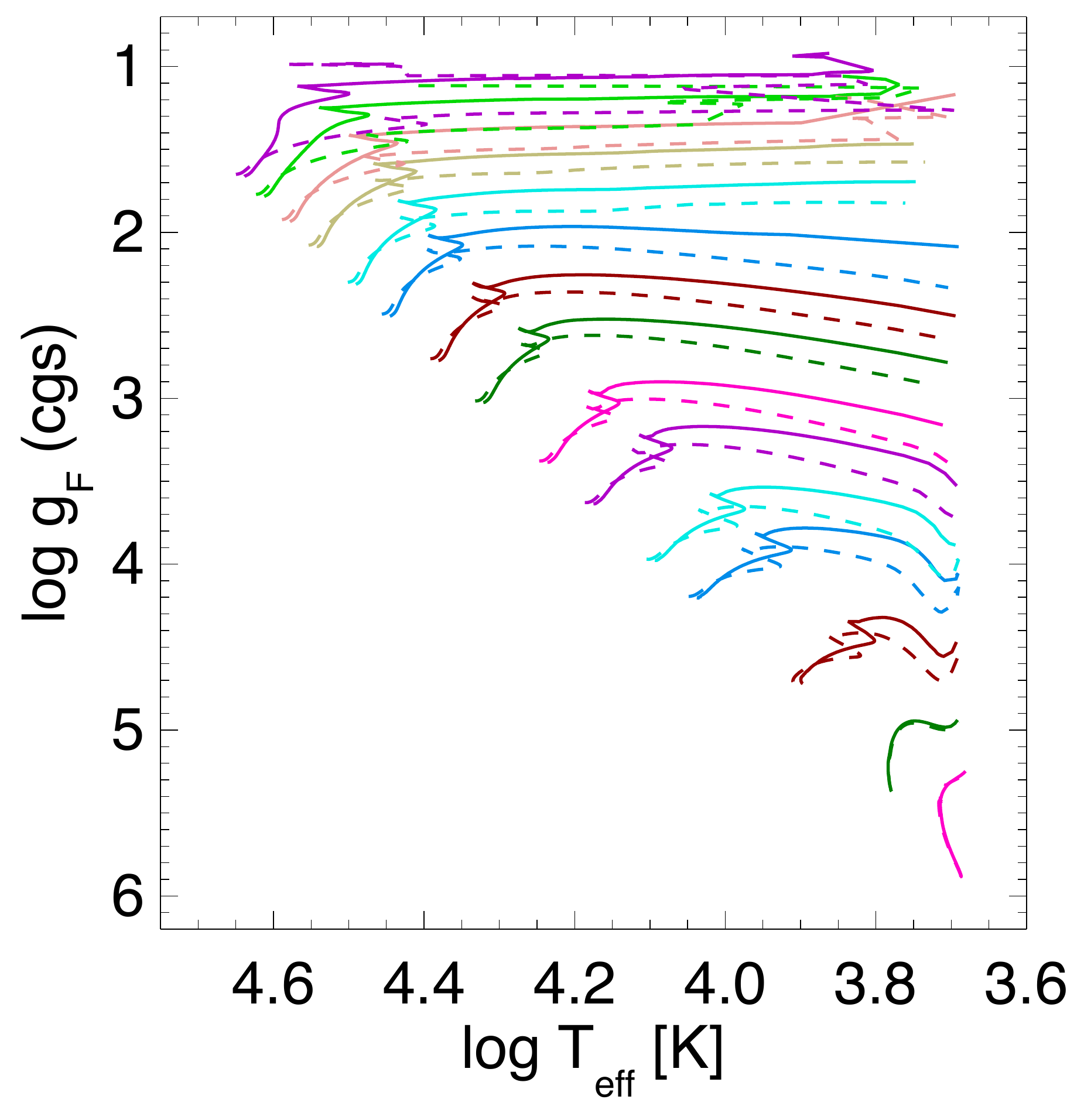}

  \end{center}
  \caption{
The evolution of stars with masses in the range from 0.8 \Msun to 40 \Msun~according to \cite{Ekstrom2012}. Upper left: Hertzsprung-Russel Diagram, upper right: mass-luminosity diagram, lower left: (\logg, log \teff)-diagram, lower right: spectroscopic HRD. Evolutionary tracks including the effects of rotation and calculated with initial rotational velocities equal to 40\% of the critical velocity are plotted solid, whereas tracks without rotation are dashed.  
} \label{figure_1}
\end{figure*}

\begin{figure}[t]
\includegraphics[scale=0.40]{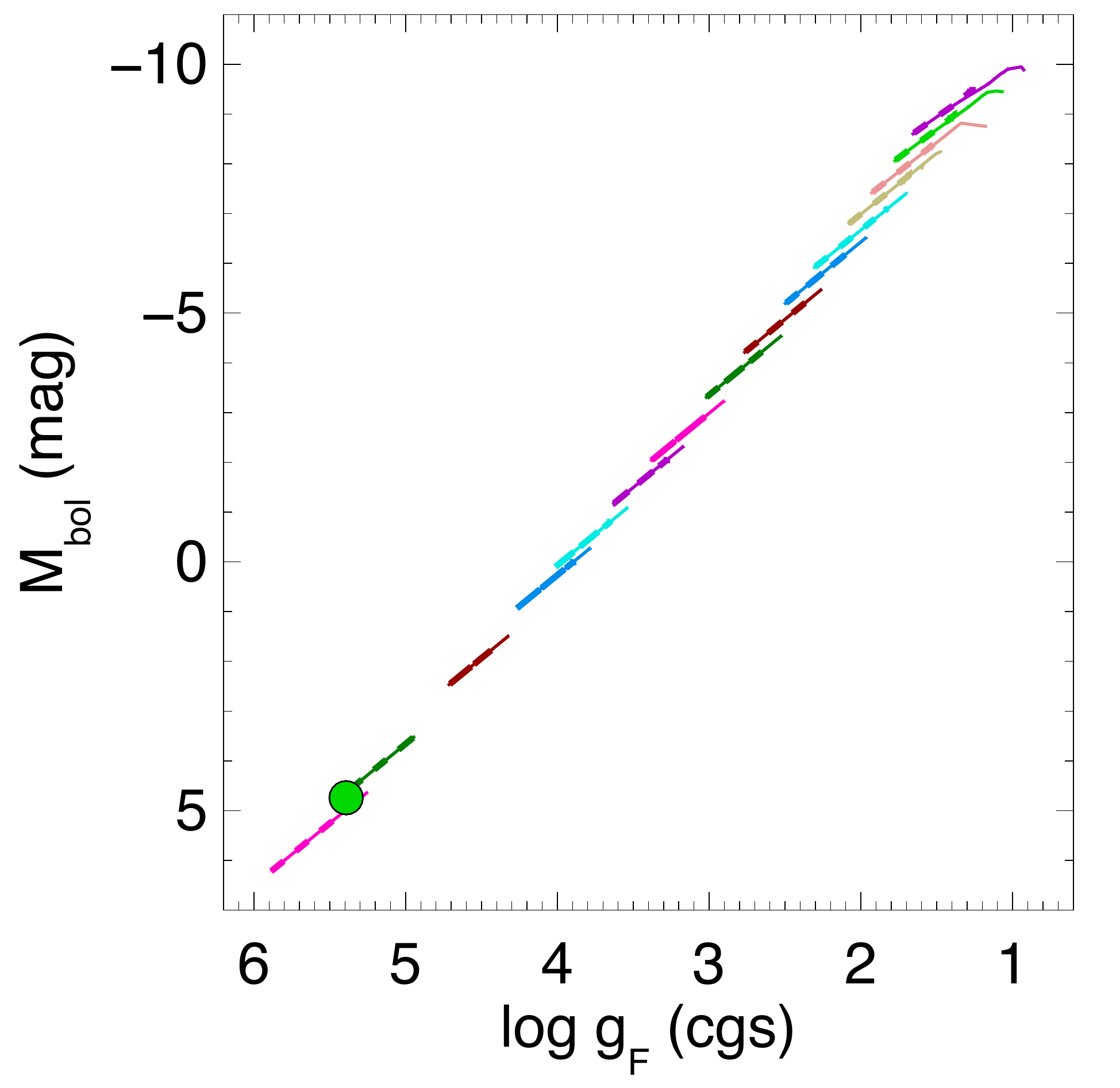}
\caption{
Stellar evolution in the (\Mbol, \loggf)-diagram. The same evolutionary tracks as in Figure \ref{figure_1} are used. They show a tight flux weighted gravity - luminosity relationship (FGLR). The location of the sun in this diagram is indicated as the green circle.
}
\label{figure_2}
\end{figure}

\begin{figure*}[ht!]
  \begin{center}
    \includegraphics[scale=0.40]{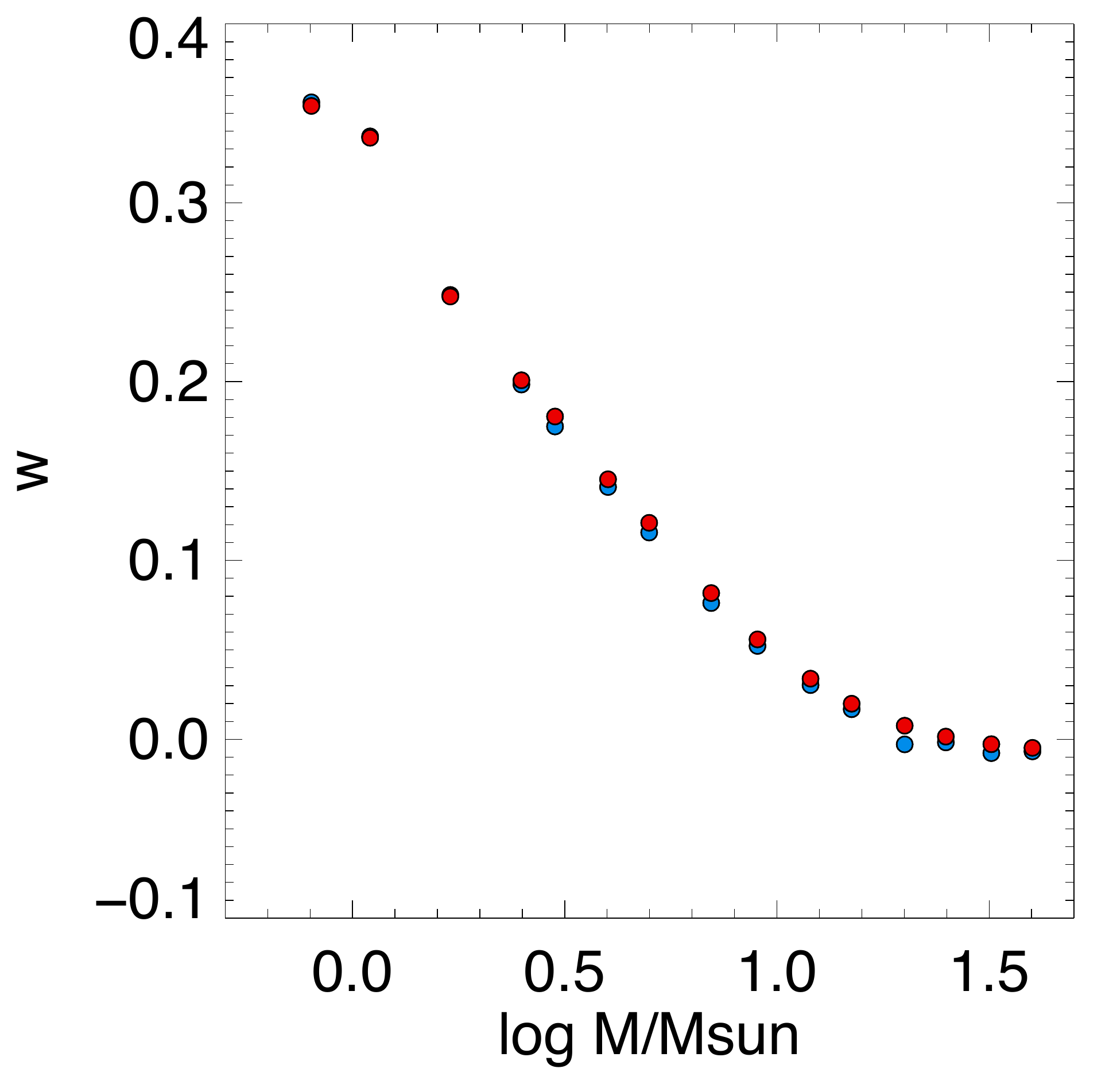}
    \includegraphics[scale=0.40]{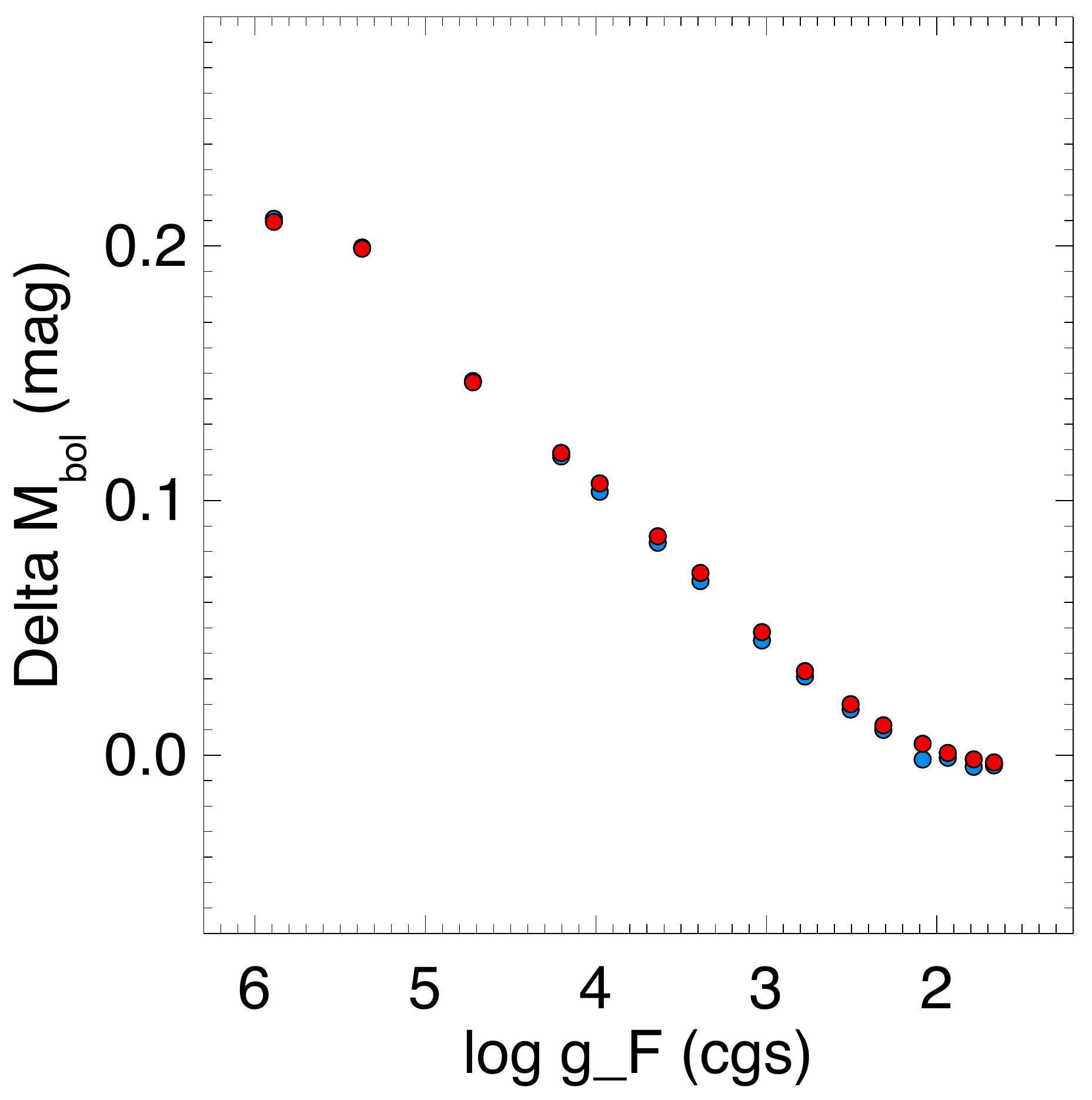}
  \end{center}
  \caption{
Effects of metallicity on the stellar evolution FGLR. Left: exponent w of equation (4) as a function of stellar mass. Right: Vertical shift of \Mbol~as a function of \loggf~caused by a change in metallicity from log Z/Z$_{\odot}$ = 0 to -0.845 (see text). Blue symbols correspond to main sequence stellar evolution models with rotation and red symbols to models without rotation.   
 } \label{figure_3}
\end{figure*}

 \begin{figure*}[ht!]
  \begin{center}
    \includegraphics[scale=0.40]{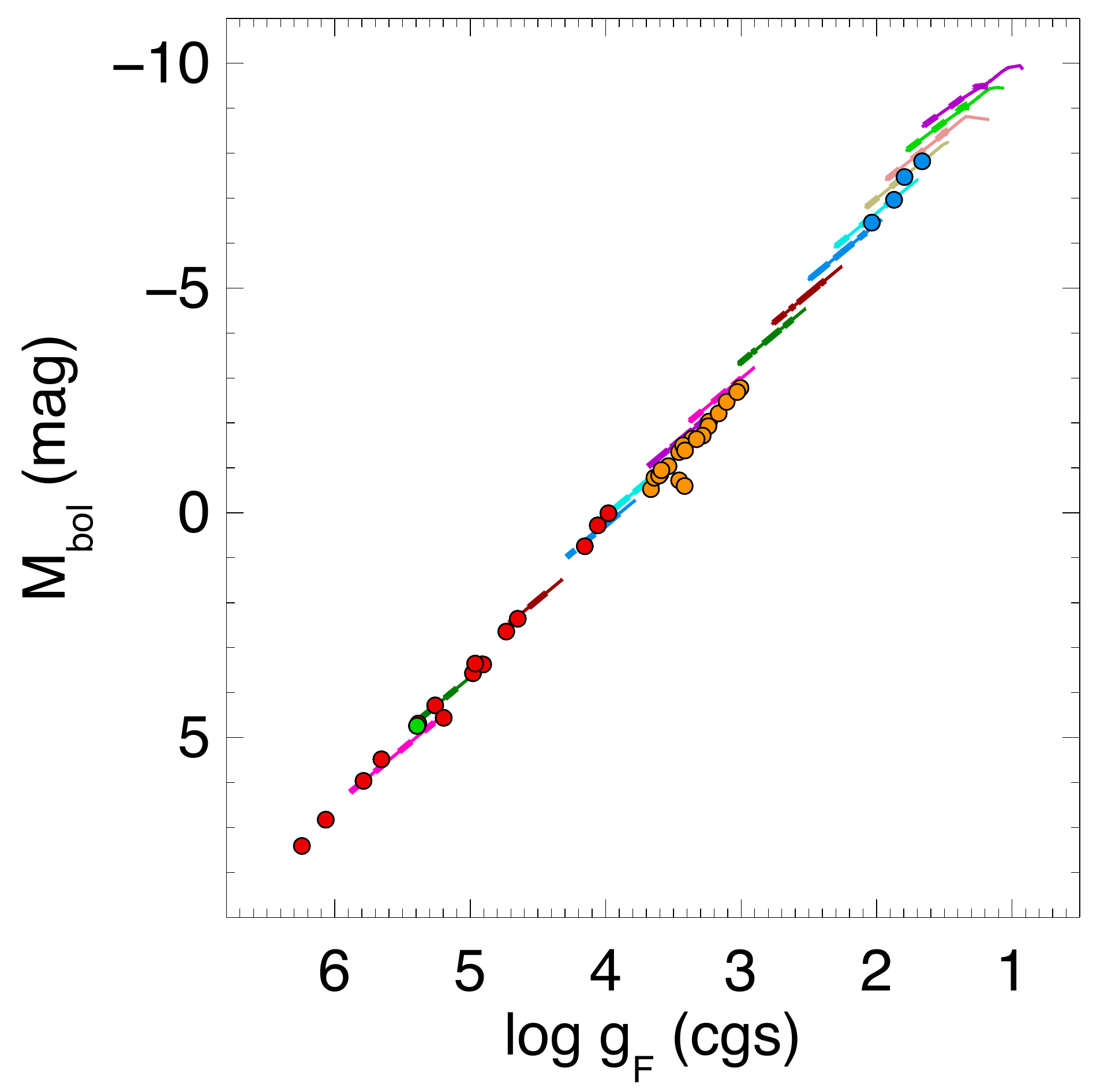}
    \includegraphics[scale=0.40]{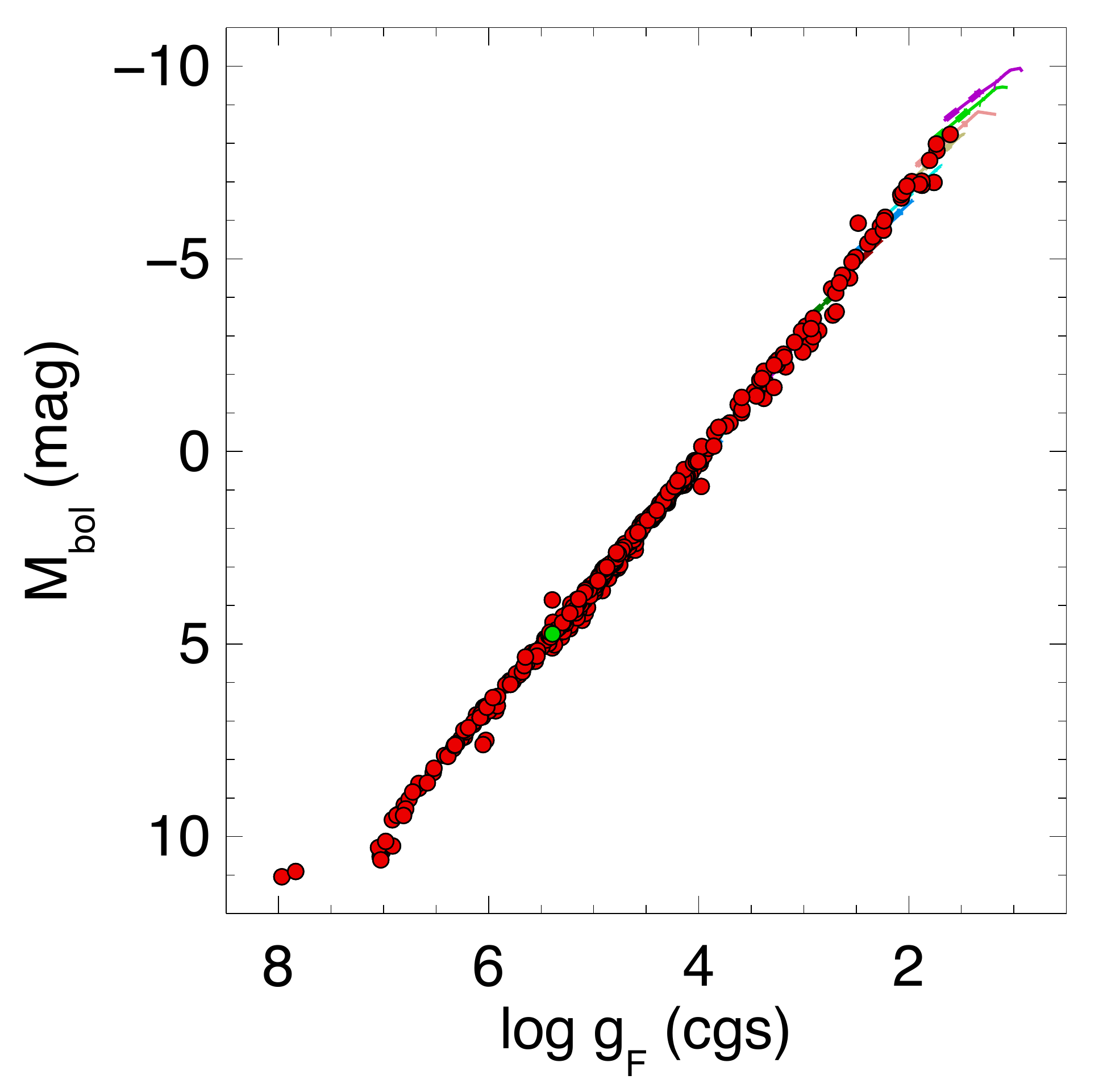}
  \end{center}
  \caption{
The stellar evolution FGLR compared with observations. Left: stars from the Gaia benchmark sample (\citealt{Heiter2015}, red), LMC detached late-type eclipsing binaries (\citealt{Graczyk2018}, orange) and LMC early-type detached eclipsing binaries (Taormina et al., 2019ab, blue). Right: detached eclipsing binaries from the DEBcat catalogue (\citealt{Southworth2015}, red). The sun is plotted in green.   
 } \label{figure_4}
\end{figure*}

\section{Stellar Evolution}

We start the introduction of the extended FGLR with a discussion of stellar evolution. For this purpose, we have chosen the grid of evolutionary tracks published by \cite{Ekstrom2012} which consist of two sets, one including the effects of stellar rotation with initial rotational velocities of 40\% of the critical velocity and the other ignoring the effects of rotation. From this grid we have selected tracks with initial masses of 0.8, 1.1, 1.7, 2.5, 3, 4, 5, 7, 9, 12, 15, 20, 25, 32, 40 \Msun, respectively, which are then displayed  in four different ways in Figure \ref{figure_1}: the classical Hertzsprung-Russell diagram (HRD), the mass-luminosity diagram, the gravity-temperature diagram, and the specroscopic Hertzsprung-Russell diagram (sHRD) as introduced by \cite{Langer2014}, where flux-weighted gravity is plotted versus effective temperature. These four diagrams summarize the well known facts, how temperature, mass and gravity are correlated with stellar brightness. Bolometric magnitude increases with stellar mass, but unfortunately, at the same stellar mass there is quite a range in magnitude, which depends on effective temperature. Stellar gravity decreases when a star evolves and becomes brighter, but the value depends on mass and temperature. The flux weighted gravity in the sHRD is a proxy of stellar luminosity and can be used to discuss stellar evolution in cases where the distance is not known. In summary, we realize that it is not straightforward to use any of these diagrams for an accurate determination of distances, in particular, when stellar rotation starts to become important.
 
However, the situation changes dramatically in Figure \ref{figure_2}, where we use the same tracks as in Figure \ref{figure_1} but plot \Mbol~as a function of \loggf. Now we obtain a very tight relationship over a large range of absolute stellar magnitudes, the 'extended FGLR'. Obviously, the prediction by stellar evolution theory is that once gravity and effective temperature are determined from the spectrum, then the flux weighted gravity can be used to estimate absolute magnitude with quite some precision.

Before we continue the discussion, we have to note two facts. First, one major reason why the relationship in Figure \ref{figure_2} is so tight is that we have extended the evolutionary tracks only until the stage before the stars start to climb the red giant branch. Had we included the RGB, then the relationship would have become wider (an explanation will be given below). Second, in our calculation of flux weighted gravity we use the effective temperature in units of $10^4$K, i.e.

\begin{equation}
  log~g_{F} = log~g - 4log {\teffeq \over 10^4 K}.
\end{equation}

In this way flux weighted gravities are of the same order of magnitude as gravities.

What is the reason why compared with Figure \ref{figure_1} the relationship between \Mbol~and \gf~becomes so tight? The answer is very simple. The flux weighted gravity is proportional to the stellar mass-luminosity ratio, \gf~$\sim$ M/L. Consequently, every individual track in Figure \ref{figure_2} (except the ones with the highest initial masses, where stellar mass-loss becomes important) forms a straight line with slope 2.5. In addition, stellar luminosity increases with mass following a power law, $L \sim M^x$.  This leads to the FGLR

\begin{equation}
 M_{\rm bol}\,= a~log~g_F +b,~~~ a = -2.5{x \over 1 - x}.
\end{equation}

With x $\sim$ 4.5 (see evolutionary tracks in Figure \ref{figure_1}) we then obtain an FGLR slope of a $\sim$ 3.2, which is very close to the slope of each individual track in the (\Mbol, \gf)-plane. This leads to a dependence of bolomectric magnitude on different mass, which coincidentally is almost aligned with the change along each individual track at constant mass. In that sense the tight FGLR is a trivial consequence of the stellar mass-luminosity relationship.

The vertical width $\Delta$\Mbol~at fixed \loggf~of an average stellar evolution FGLR constructed from Figure \ref{figure_2} can be estimated as

\begin{equation}
 \Delta M_{\rm bol}\,= \pm {1 \over 2} (a - 2.5)~\Delta log~g_F,
\end{equation}

where $\Delta$\loggf~is the average length in flux weighted gravity of an individual track of fixed solar mass in Figure \ref{figure_2}. For 
$\Delta$\loggf $\sim$ 0.55 and a = 3.2 we obtain $\Delta$\Mbol = $\pm$0.2~mag, which is an encouragingly small number.

As is well know but (for simplicity) not shown in Figure \ref{figure_1}, stars increase their luminosity once they reach the red giant branch at low temperatures. This gain in luminosity is particularly pronounced at lower stellar masses. For instance, a track with 1.1 \Msun~(the lower straight green line in Figure \ref{figure_2}) would easily evolve (with approximately constant mass) up to \Mbol~$\sim$ -3.5~mag at \loggf~$\sim$ 2.1. Thus, as quantitatively described by equation (3), including RGB-stars has the potential to make the extended FGLR much wider. This is the reason, why we exclude RGB stars from the relationship.

As is well known, stellar luminosities depend on stellar metallicity Z. This will have an effect on the stellar evolution FGLR shown in Figure \ref{figure_2}, which is based on evolutionary tracks calculated for solar metallicity Z$_{\odot}$. For main sequence stars the shift in luminosity caused by a change in Z is usually described by

\begin{equation}
  L \sim ({Z \over Z_{\odot}})^{-w}M^x,
\end{equation}

where w is a positive number. As a consequence of equation (4) and because of \gf~$\sim$ M/L we expect a shift at fixed stellar mass to lower \Mbol~and lower \loggf~in Figure \ref{figure_2}, when the metallicity is reduced. As a result, the new extended FGLR at lower metallicity has the same form as equation (2) and the change in bolometric magnitude at a fixed value of \loggf~can be described by

\begin{equation}
  \Delta M_{\rm bol}(Z)\,= -w(a - 2.5)log{Z \over Z_{\odot}}.
\end{equation}

Obviously, the crucial parameter to estimate the influence of metallicity on the extended FGLR is the exponent w. We can use the stellar evolution calculations by \cite{Georgy2013} to determine w. These calculations use exactly the same physics input as \cite{Ekstrom2012} but adopt a significantly lower metallicity log Z/Z$_{\odot}$ = -0.845. Figure \ref{figure_3} shows that w is a function of stellar mass. At low stellar mass we find w $\sim$ 0.35 but for higher mass w quickly approaches zero. The corresponding changes in \Mbol~calculated with equation (5) are also shown in Figure \ref{figure_3}. At lower mass and high \loggf~ the change is 0.2 mag, which corresponds to the half width of the stellar evolution FGLR at solar metallicity as discussed above and described by equation (3). For higher masses the effects are negligible. We conclude that for large differences from solar metallicity the extended FGLR may show an effect at the high flux weighted gravity end.

\section{Comparison with Observations}

\begin{figure}[t]
\includegraphics[scale=0.40]{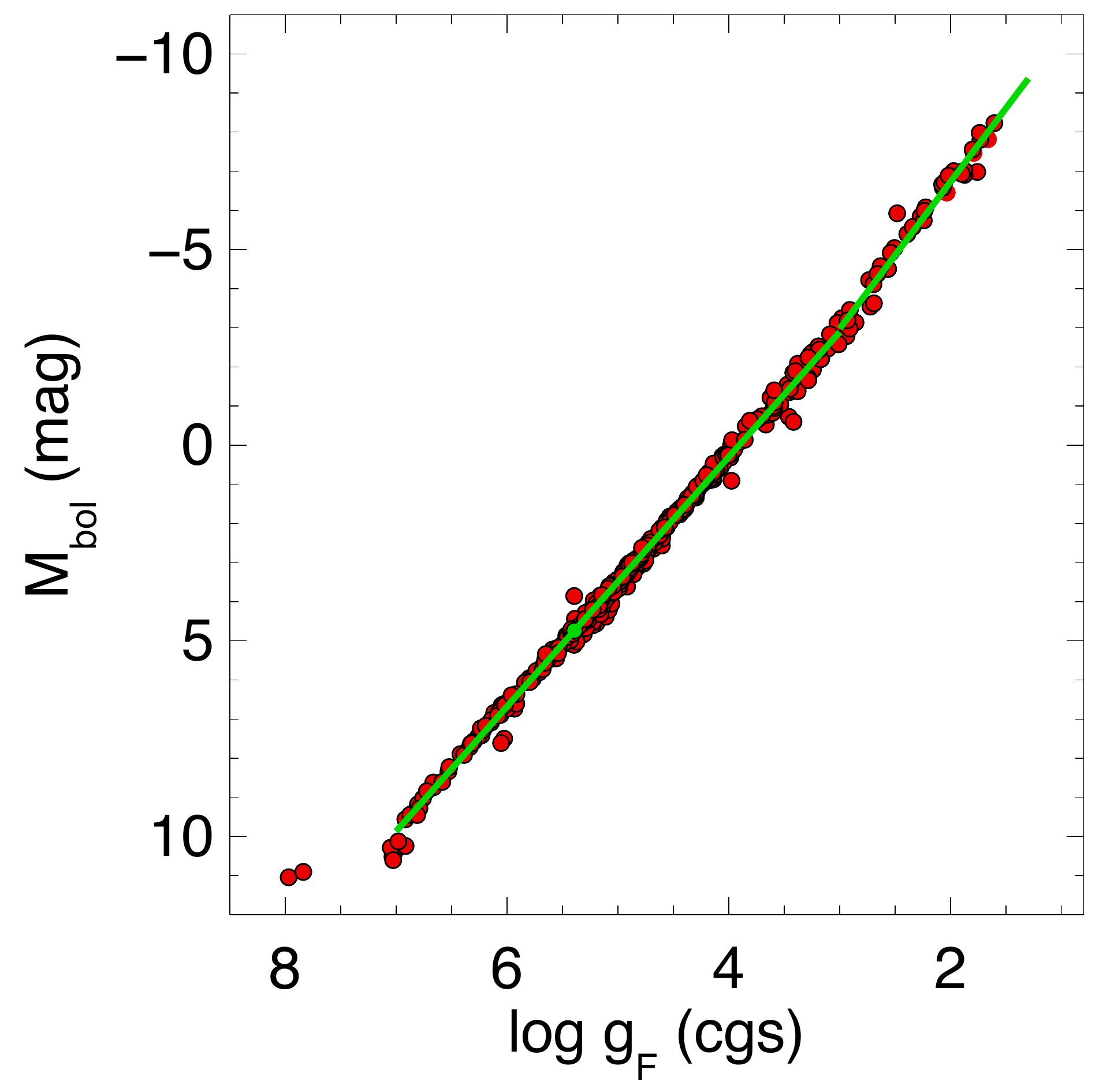}
\caption{
FGLR linear regression fit to the combined data of Figure \ref{figure_4} (see text).
}
\label{figure_5}
\end{figure}

\begin{figure*}[ht!]
  \begin{center}
    \includegraphics[scale=0.40]{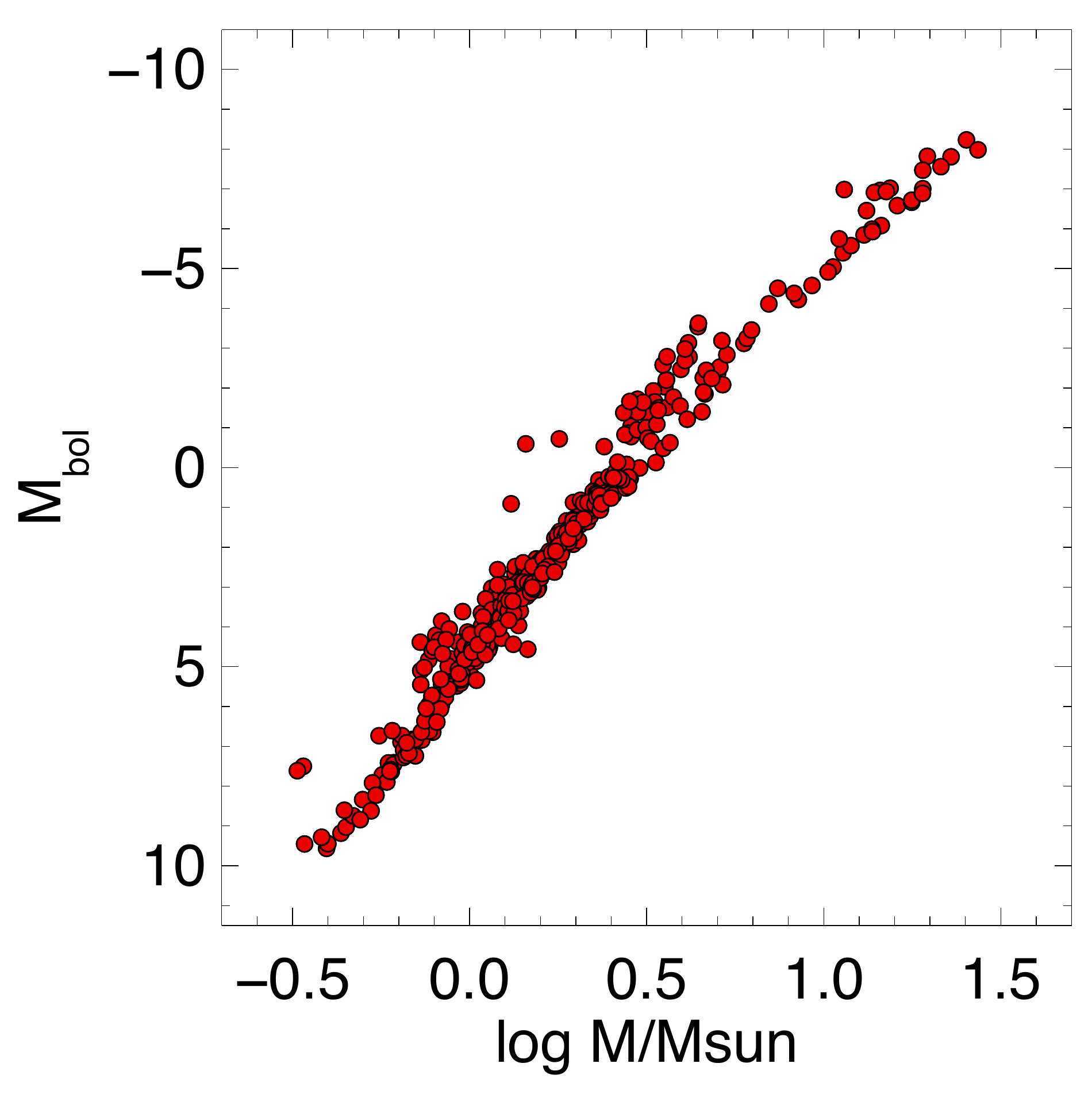}
    \includegraphics[scale=0.40]{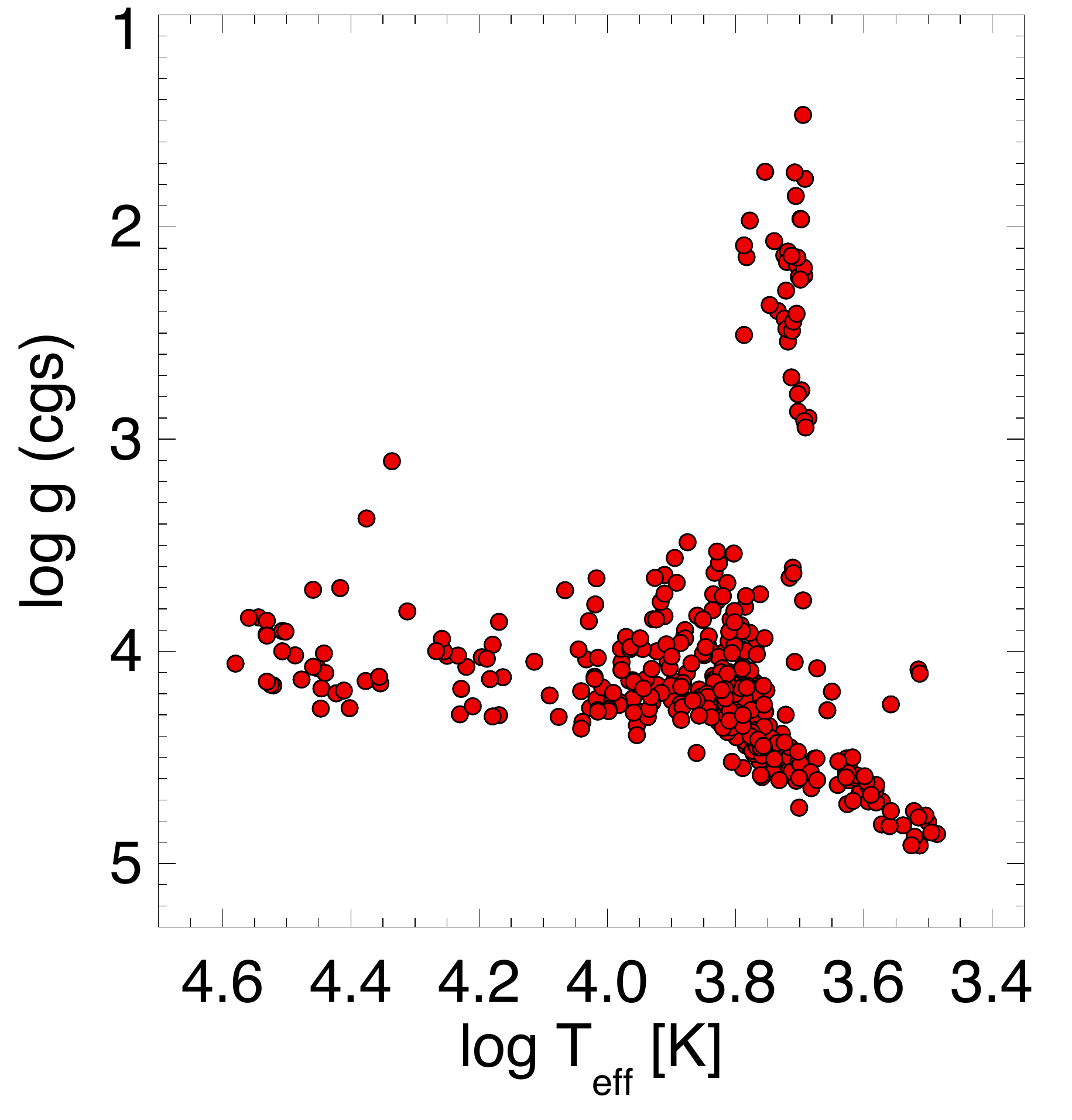}
  \end{center}
  \caption{
Physical properties of the stars used for the regression in Figure \ref{figure_5}. Left: masses and luminosities, right: gravities and effective temperatures.      
 } \label{figure_6}
\end{figure*}

\begin{figure*}[ht!]
  \begin{center}
    \includegraphics[scale=0.40]{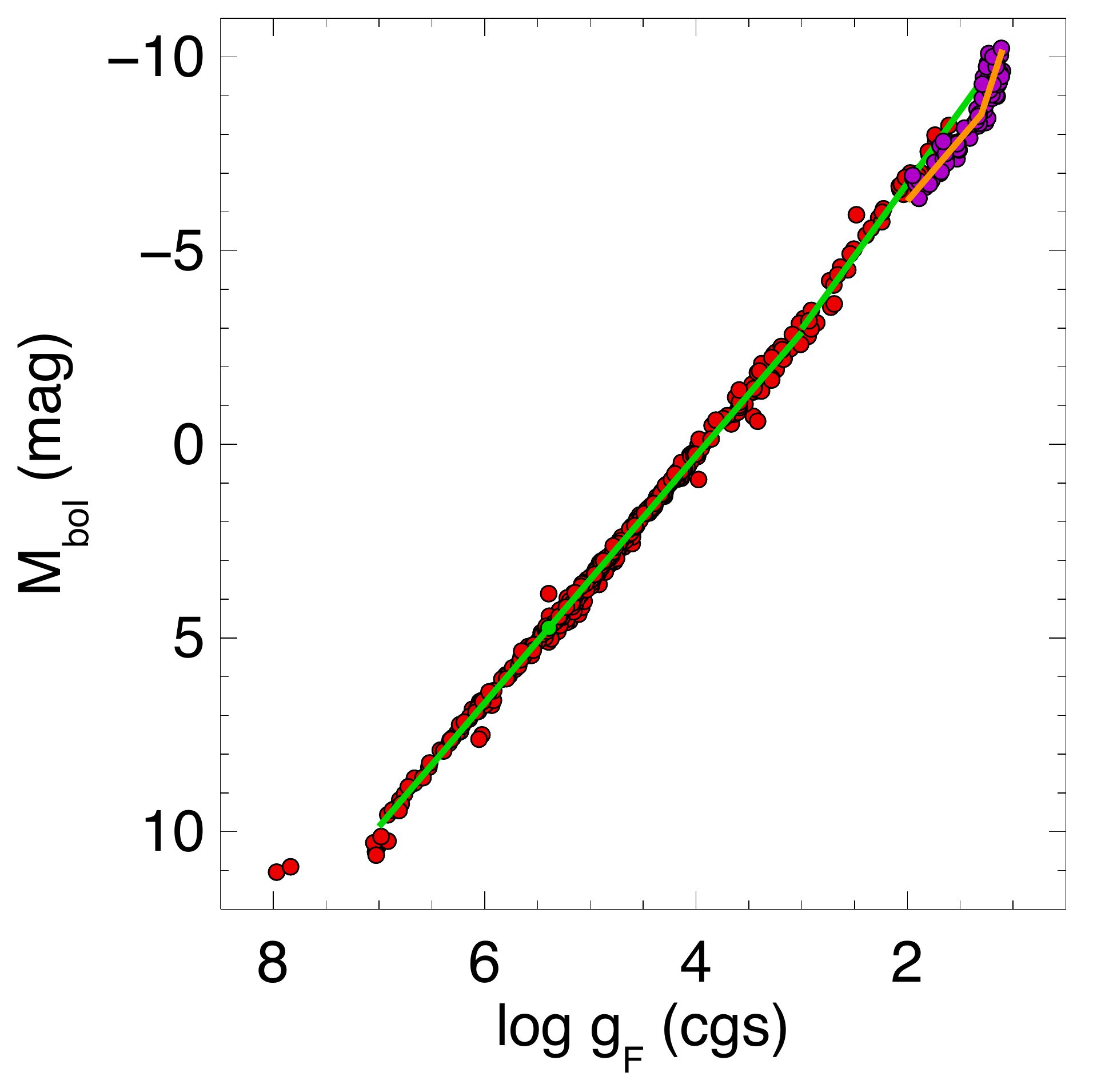}
    \includegraphics[scale=0.40]{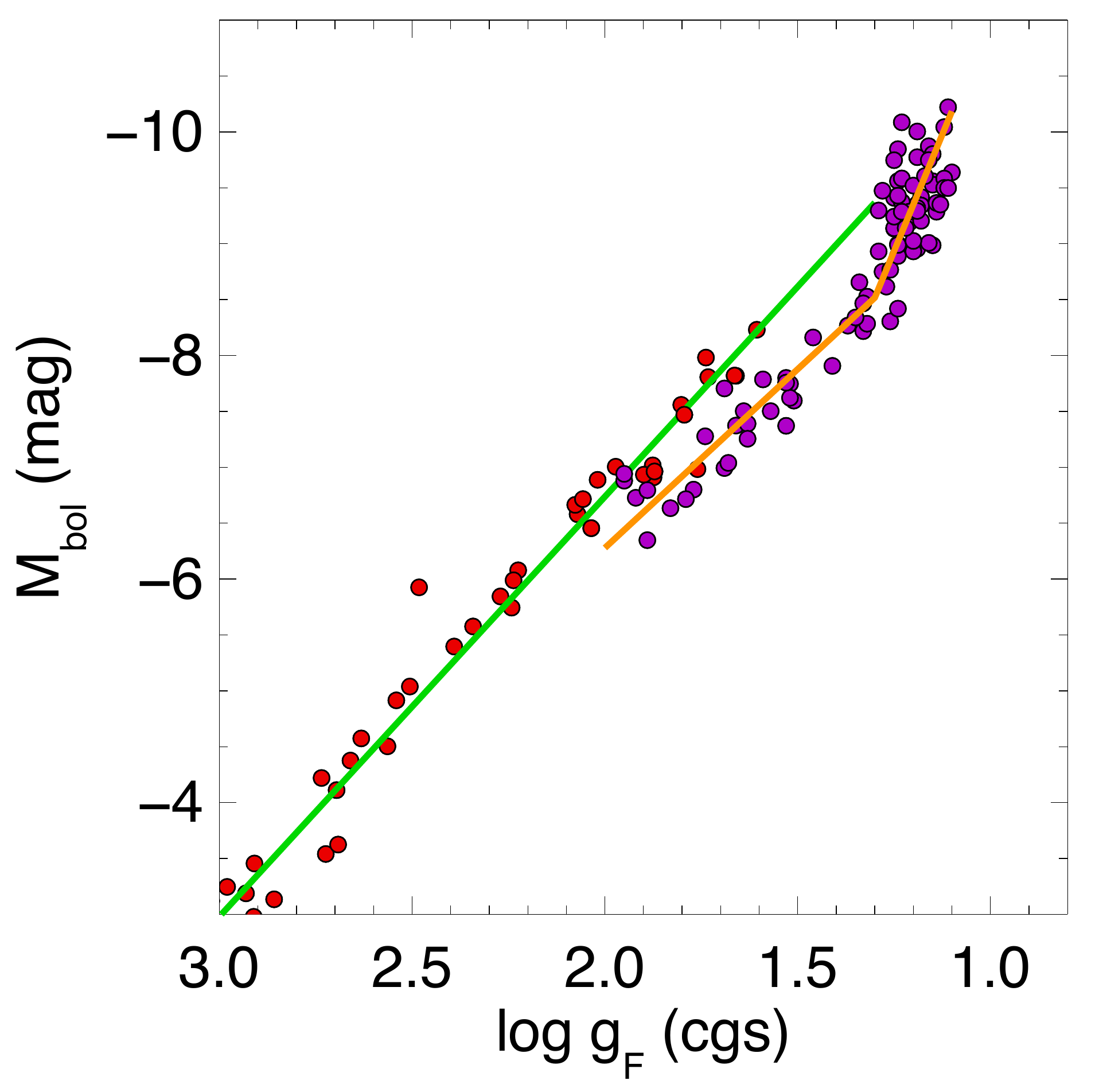}
  \end{center}
  \caption{
The FGLR of Figure \ref{figure_5} compared the FGLR of blue supergiant stars (violet with orange regression) as obtained from the work by \cite{Urbaneja2017}. Left: full range in \loggf, right: restriction to low values of \loggf.      
 } \label{figure_7}
\end{figure*}

The predicted relationship of Figure \ref{figure_2} needs to be compared with observations. For this purpose we collated samples with accurate stellar parameters either obtained from quantitative spectroscopic analysis or binarity or parallaxes or a combination of all three. In a first step we use the Gaia benchmark stars by \cite{Heiter2015}, the LMC detached late-type eclipsing binaries by \cite{Graczyk2018} and the detached LMC early-type eclipsing binaries studied \cite{Taormina2019} and Taormina et al., 2019b (submitted to ApJ). To be consistent with the previous section we have excluded stars on the red giant branch by eliminating all objects with \teff~$\le$~4900K and with \logg~$\le$~3.5. We have also avoided objects with extremely low metallicity [Fe/H] $\le$ -2.0. Figure \ref{figure_4} shows the FGLR comparison of these objects with the prediction by stellar evolution. The agreement is striking. 
In a next step we use as an additional sample the detached eclipsing binaries compiled in the DEBcat catalogue by \cite{Southworth2015}. Again we have eliminated all objects on the red giant branch. In addition, in order to avoid double counting with the previous sample we also excluded the objects from \cite{Graczyk2018} and \cite{Taormina2019}, which are also contained in DEBcat. We also excluded the pre-main sequence binary system EPIC 203710387. The comparison with this sample is also shown in Figure \ref{figure_4}. Again we find excellent agreement with stellar evolution theory.

Figure \ref{figure_5} combines the two stellar samples of Figure \ref{figure_4} and shows a linear regression fit. Since the data indicate a change of the slope at \loggf~$\sim$ 3.0 caused by a change in the exponent x of the stellar mass-luminosity relationship (see discussion in the previous section), we use a two component fit. We introduce log g$_{F}^{b}$ = 3.0 and the flux weighted gravity of the sun log g$_{F}^{\odot}$ = 5.39 and fit two linear regression to the data in the range 7.0 $\geq$ log g$_F$ $\geq$ log g$_{F}^{b}$ and log g$_F \le$  log g$_{F}^{b}$, respectively:

\begin{equation}
 \mbox{\lgf}\, \geq \,\log\,g_{\!\mbox{\scriptsize \,\sc f}}^{\!\mbox{\scriptsize \,b}}:\,   M_{\rm bol}\,=\,a (\mbox{\lgf}\,-\,log~g_{F}^{\odot})\,+\,b
\end{equation}
and 
\begin{equation}
\mbox{\lgf} \leq \,\log\,g_{\!\mbox{\scriptsize \,\sc f}}^{\!\mbox{\scriptsize \,b}} \,:\,     M_{\rm bol}\,=\,a_{\rm l} (\mbox{\lgf}\,-\,\log\,g_{\!\mbox{\scriptsize \,\sc f}}^{\!\mbox{\scriptsize \,b}})\,+\,b_{\rm l}  .
\end{equation}

We obtain a = 3.19$\pm0.01$, b = 4.74$\pm0.01$ with a scatter of $\sigma$ = 0.17 mag for the high gravity range. For the low gravity domain the fit yields a$_{\rm l}$ = 3.76$\pm0.11$, b$_{\rm l}$ = -2.98$\pm0.09$ with a scatter of $\sigma_{\rm l}$ = 0.29 mag. The scatter is in agreement with the expectation from Section 2 and equation (3).

We note that 23 of the 445 stars used for the regression have metallicities log Z/Z$_{\odot}$ lower than -0.6 down to the most extreme value of -1.80. The average metallicity of these 23 objects is about -1.0. We find that all these objects except one have \Mbol~values larger than given by the regression of equations (6) and (7) with an average $\Delta$\Mbol $\sim$ 0.3 mag. This confirms the conclusion of the previous section that extreme changes in metallicity may have an effect on the etxtended FGLR.

Figure \ref{figure_6} summarizes the physical properties of the stars included in the regression. We cover a mass range from 0.3 to 20 \Msun. The data reveal a change in slope of the mass-luminosity relationship which correpondes to the change in slope of the FGLR relationship of Figure \ref{figure_5}. We also note that while we encounter quite a range of bolometric magnitudes at a fixed mass the scatter in the FGLR is much smaller. This is explained by the behaviour of the evolutionary tracks in Figure \ref{figure_2} as discussed in the previous section.

Figure \ref{figure_6} also shows the gravities of the total sample used for the regression. While a large fraction of the objects is still on the main sequence, a significant amount has left the mains sequence and is evolving towards the red giant phase. In consequence, the extended FGLR of Figure \ref{figure_5} and described by equations (6) and (7) holds for stars between  0.3 to 20 \Msun~in all phases from the zero age main sequence to the beginning of the red giant branch. There is, however, one caveat at the high mass end, which is dicussed in the next section.

\section{Massive blue supergiant stars}

%\begin{figure*}[ht!]
\begin{figure}[]  
  \begin{center}
    \includegraphics[scale=0.45]{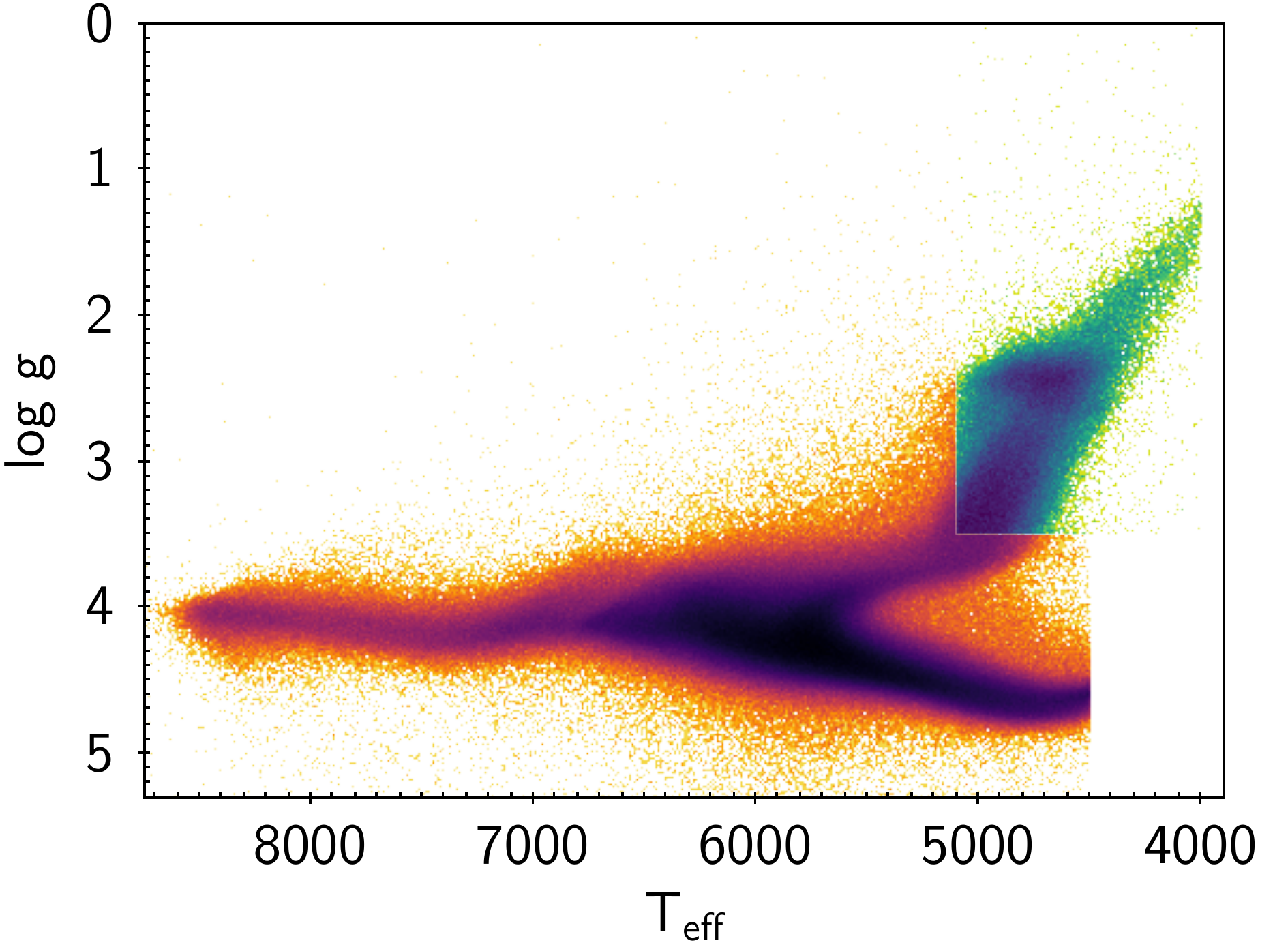}
  \end{center}
  \caption{
Stellar effective temperatures and gravities of the LAMOST sample described in the text: (\logg,~\teff) - diagram. The density of the non-RGB stars (to be compared with the extended FGLR) is shown in orange tones and that of the RGB stars in green.
 } \label{figure_8}
\end{figure}
%\end{figure*}

\begin{figure*}[ht!]
  \begin{center}
    \includegraphics[scale=0.375]{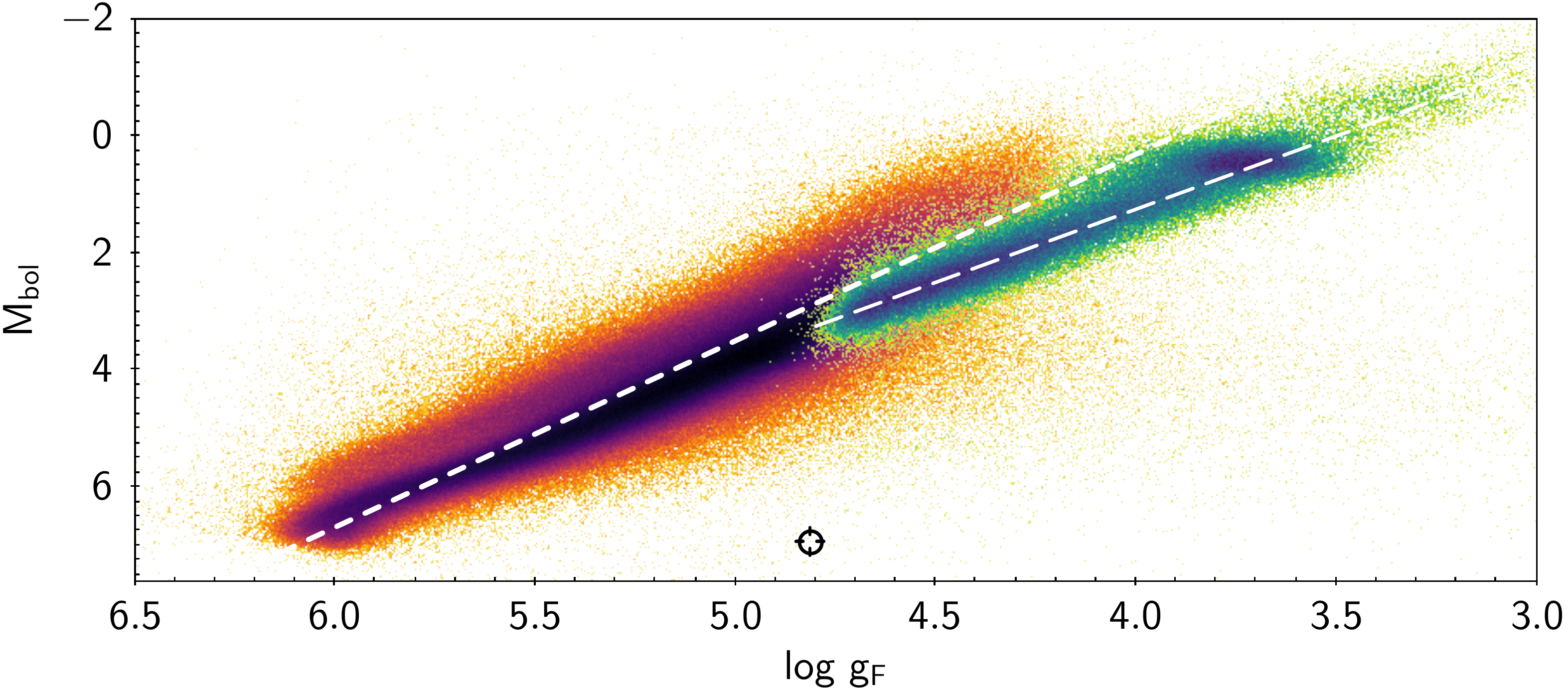}
    \includegraphics[scale=0.30]{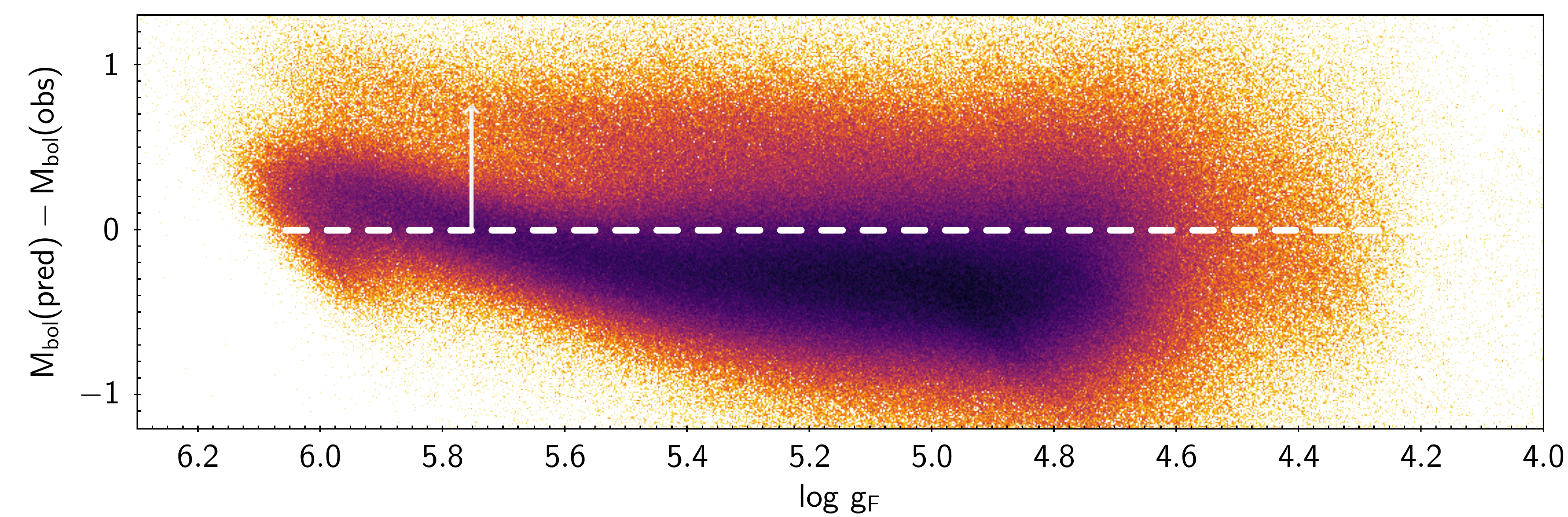}
  \end{center}
  \caption{
Flux weighted gravities and absolute bolometric magnitudes of the LAMOST sample. Top: FGLR diagram of the same objects as in Figure \ref{figure_8}, non-RGB stars in orange, RGB stars in green. The bold dashed line corresponds to the FGLR of equation (6) for non-RGB stars. The long dashed line along the FGLR sequence of the RGB stars corresponds to a stellar evolutionary track with 1.1~\Msun, which has a slope of 2.5, expected for objects of similar mass merely varying in \Mbol. For discussion see text.; bottom: difference $\Delta$\Mbol~from the predicted relation of equation (6) as a function of \loggf~for the non-RGB stars. Besides the extended FGLR, the 'binary sequence' of unresolved binaries of comparable luminosity is also apparent and indicated by the arrow of 0.75~mag length.
 } \label{figure_9}
\end{figure*}

The FGLR was originally introduced by \cite{Kudritzki2003, Kudritzki2008} for massive blue supergiant stars (BSG) in the effective temperature range between 8000 to 25000K. Because of their extreme brightness at visual light these objects are ideal spectroscopic extragalactic distance indicators and, consequently, the FGLR technique using BSG has been applied to determine distances to a considerable amount of galaxies (see \citealt{Kudritzki2016} and references, therein). More recently, \cite{Urbaneja2017} have re-calibrated the BSG FGLR using a large sample of objects in the LMC. Figure \ref{figure_7} shows the comparison of the result from Figure \ref{figure_5} with the BSG from \cite{Urbaneja2017}. The offset is caused by severe mass-loss when massive stars start to evolve away from the main sequence (see \citealt{Meynet2015}). All the BSG in Figure \ref{figure_7} have significantly lower gravities than the red objects of comparable \Mbol~in the figure. This means that below flux weighted gravities of \loggf~= 2.0 and for effective temperatures below \teff~= 25000K one needs to check whether stellar gravities \logg~are above or below the value of 3.0 in order to distinguish between the two FGLR branches.

\section{The extended FGLR for two million stars with LAMOST spectroscopy}

In Section 3 we have used stellar effective temperatures and gravities of very high precision to confirm the existence of the extended FGLR. In this section we investigate whether this relationship can be recovered, when data of lower precision are used, for instance, obtained from 'mass production' spectroscopy of a large sample of medium or low resolution spectra and of moderate S/N ratio. 

At present, most stellar spectra come from a few of the large spectroscopic surveys (SDSS/APOGEE, \citealt{Majewski2017}; GALAH, \citealt{Buder2018}; LAMOST, \citealt{Luo2015a,Luo2015b}). In particular, the LAMOST survey provides spectra for over six million stars. These spectra have a resolution of only R $\sim$ 1800, but are good enough to yield element abundances, and temperatures precise to typically ~70K and gravities \logg ~precise to typically 0.06 \citep{Xiang2019}. If one focusses on stars before the RGB phase as in the previous sections, the LAMOST survey stars should of course also follow the extended FGLR. This provides a test of this relation.

While LAMOST samples stars across a large range of effective temperatures, the derived (\teff,~\logg) values have only been validated for 4000K $\le$ \teff $\le$ 8500K. We now construct a FGLR diagram, drawing on the LAMOST DR5 sample from \cite{Xiang2019} after making the following cuts: to avoid effects of metallicity as discussed in the previous sections we include only objects with [Fe/H] $\ge$ -0.6. For stars with \logg~$\le$ 3.5 we select only those with \teff~$\ge$ 5100K, to eliminate giants as in section 3. We have chosen a higher temperature value than in section 3, because for the LAMOST sample the red giant branch seems to extend to somewhat higher \teff~values (see Figure \ref{figure_8}). For stars with \logg~$\ge$ 3.5 we restrict Teff to 4500K $\le$ \teff~$\le$ 8500K, as the \logg~values of cooler main sequence stars are known to have significant systematic problems. We further restrict the sample to spectra with S/N $\ge$ 20, and good Gaia DR2 parallax estimates with parallax/parallax error $\ge$ 20. These selection criteria leave 2.19 million objects, which occupy the (\logg,~\teff)-diagram, as shown in Figure \ref{figure_8}.

For all of these objects we construct the observed absolute bolometric
magnitude via

\begin{equation}
  M^{obs}_{bol} = m_{W1} + 5\log{\varpi} - 10 + BC_{W1},
  \end{equation}
    
where $m_{W1}$ is the dust-reddening insensitive apparent magnitude in the WISE band 1 and $\varpi$ is the parallax measured in milli-arcseconds. $BC_{W1}$ is the bolometric correction, which we derive from Padova isochrones, which imply the following approximation

\begin{equation}
  BC_{W1} = 7.5(\log{T_{eff}} - 3.75) -1.55.
\end{equation}

This approximation is good to 0.02~mag over the relevant temperature range.

With an effective temperature uncertainty of $\sim$70K (see above) or 1.6 to 1.0 \% the uncertainty of the bolometric correction is $\sim$ 0.05 mag. Therefore, depending on the uncertainty of the Gaia parallaxes we expect uncertainties $\Delta$$M^{pred}_{bol}$ between 0.12 to 0.05 mag for the observed absolute bolometric magnitudes (we have neglected the photometric uncertainties of W1 Wise magnitudes in this estimate). As we know from cross-validation \citep{Xiang2019} the scatter in the \logg~estimates from LAMOST is 0.06 dex. This error dominates the uncertainty of \loggf~which has the same value.

If we the compare the $M^{pred}_{bol}$ predicted by the extended FGLR of equation (4) to $M^{obs}_{bol} $, we find
remarkable agreement, as Figure \ref{figure_9} demonstrates. The 1$\sigma$-scatter is $\sim$ 0.18~mag at the low luminosity end and $\sim$ 0.27~mag at high luminosities with a small offset of $\sim$ 0.2~mag. Also, the binary caustic, a sequence of unresolved binaries of similar luminosities, is clearly apparent with the expected 0.75~mag (factor of two in luminosity) offset. While a detailed comparison with recent studies of the binary population is beyond the scope of this paper, we note that the observed strength of the binary sequence in the LAMOST data seems to be in qualitative agreement with \citealt{El-Badry2018} (see their Figure 9) and \citealt{El-Badry2019} (see again Figure 9). 

The error $\Delta$\loggf~= 0.06 translates into a $3.19\times 0.06 = 0.19$~mag scatter in $M^{pred}_{bol}$. Therefore, most of the scatter is expected to result from the \logg~uncertainties combined with the intrinsic width of the FGLR as discussed in Section 2. The systematic offset of 0.2 mag could be easily explained, if the \logg~values from the low-resolution LAMOST spectra had a systematic offset of only $\sim$0.06 dex. Therefore, this constitues a remarkable affirmation of the extended FGLR, even on the scale of massive spectroscopic surveys. Note that a significant fraction of the remaining overall LAMOST sample, has parallax-based distance uncertainties $\ge$ 10\%. For those objects the extended FGLR can provide distances potentially better than Gaia parallaxes making this a competive distance predictor even for a vast sample with stellar parameters \teff~and \logg~derived on an industrial scale.

The LAMOST RGB stars plotted in green in the (\logg,~\teff)-diagram are also shown in Figure \ref{figure_9}. As we can see, these objects also form a FGLR sequence, however with a slope of $\sim$2.5 corresponding to a stellar evolution track with a constant stellar mass as discussed in section 2. Indeed, if we overplot the RGB extension of the evolutionary track with 1.1~\Msun~shown in Figure \ref{figure_1}, we find a good representation of the observed RGB FGLR sequence. It seems that the tight relation for RGB stars in LAMOST arises from the fact that the large majority of stars is 'old' ($\ge$2Gyrs) with masses between 1 and 2 \Msun.

\section{Discussion}

We have shown that for stars between 0.3 to 20 \Msun~and in evolutionary phases from the main sequence to the beginning of the red giant branch the flux weigthed gravity \gf~is tightly correlated with absolute bolometric magnitude. The FGLR correlation holds over 17 stellar magnitudes from \mbol~= 9.0 mag to -8.0 mag with a scatter of $\sigma$ = 0.17 mag below \Mbol~= -3.0 and $\sigma$ = 0.29 mag above this value. Thus, with effective temperatures and gravities derived from stellar spectrum applying the standard methods of model atmosphere analysis it is possible to estimate stellar luminosities L with a precision of 15 to 30 \%. Our approach here differs from other recent approaches to determine spectrophotometric distances or luminosities \citep{Coronado2018, Hogg2019, Anders2019} in that it is based on a very simple relation that directly results from elementary stellar physics. For many astrophysical purposes this a valuable tool. For instance, in combination with multi-color photometry allowing for a simultaneous determination of interstellar reddening this will allow to determine distances with a precision of 7 to 15 \%. The comparison with a sub-sample of the LAMOST survey consisting of 2.2 million stars, for which \teff and \logg~where obtained from an automated analysis of low resolution spectra, confirms these findings. Thus, the extended FGLR provides the means for a competetive distance determination method, as soon as the precision of parallax measurements is less than $\sim$ 10\%. Alternatively, for large samples of stars, where very precise distances are available, the comparison of observed with predicted absolute magnitudes can be used to construct accurate 3D-maps of interstellar extinction. 

\acknowledgments
This work was initiated during the Munich Institute for Astro and Particle Physics (MIAPP) 2018 program on the extragalactic distance scale. RPK gratefully acknowledges support by the Munich Excellence Cluster Origins Funded by the Deutsche Forschungsgemeinschaft (DFG, German Research Foundation) under Germany's Excellence Strategy EXC-2094 390783311. We thank our colleagues Travis Berger, Andreas Burkert, Joachim Puls and Eva Sextl for critical and stimulating discussion along this project. \\

\pagebreak
\bibliography{ms}

\end{document}